\documentclass[a4paper,fleqn]{cas-dc}

\usepackage[numbers]{natbib}

\begin{document}

\let\WriteBookmarks\relax
\def\floatpagepagefraction{1}
\def\textpagefraction{.001}
\shorttitle{Determining the $\Xi$-Nucleus Potential}
\shortauthors{P. T. Nyein et~al.}


\title [mode = title]{Determining the \texorpdfstring{$\Xi$}{Xi}-Nucleus Potential from the Measured Binding Energies of \texorpdfstring{$^{15}_{\Xi^-}$C}{15-Xi-C} }     





\author[1]{Paing Thit Nyein}[
type=editor,
]
\ead{paingthitnyein.ptn@gmail.com}


\affiliation[1]{organization={Department of Physics, Defense Services Academy},
                addressline={Pyin Oo Lwin}, 
                city={Mandalay},
                country={Myanmar}}

\author[1]{Kyaw Kyaw Naing}[
]
\ead{drkyawkyawnaing4550@gmail.com}
\author[2]{Htun Htun Oo}
\ead{htunhtun.oo93@gmail.com}
\affiliation[2]{organization={Department of Physics, Panglong University},
            addressline={Southern Shan State}, 
            city={Panglong},
            country={Myanmar}}

\author[3]{Masahiro Yamaguchi}[]
\ead{yamagu@rcnp.osaka-u.ac.jp}

\author[3,4]{Hiroyuki Kamada}[ orcid=0000-0001-6519-9645]
\ead{kamada@rcnp.osaka-u.ac.jp}

\affiliation[3]{organization={Research Center for Nuclear Physics (RCNP), Osaka University},
            addressline={Ibaraki}, 
            city={Osaka},
            postcode={567-0047}, 
            country={Japan}}
\affiliation[4]{organization={Department of Physics, Faculty of Engineering, Kyushu Institute of Technology},
            addressline={Tobata}, 
            city={Kitakyushu},
            postcode={804-8550}, 
            country={Japan}}     



\begin{abstract}
The recent observation of the deeply bound $\Xi$ - hypernucleus $^{15}_{\Xi^-}$C through the IRRAWADDY and KINKA events provided a crucial benchmark for determining the $\Xi$-nucleus interaction. This work aims to constrain the depth of this potential by calculating the binding energy $B_{\Xi}$ of the ${}^{15}_{\Xi^-}\mathrm{C}$ system, which forms a $\Xi^{-}$-${}^{14}\mathrm{N}$ bound state. We achieve this by numerically solving the {Schr\"odinger equation} for a $\Xi$ hyperon within a phenomenological {Woods-Saxon potential}, using the stable {Numerov method}, incorporating the {Coulomb interaction}. For a potential well depth $V_0 = 12\ \mathrm{MeV}$, our calculations yield a ground state $ J^\pi_n=0^+_1$ binding energy of $6.35\ \mathrm{MeV}$ and a excited state $1^-_1$ energy of $0.87\ \mathrm{MeV}$. These results are in excellent agreement with the ground state $0^+_1$ of the  IRRAWADDY event ($B_{\Xi^-} = 6.27 \pm 0.27\ \mathrm{MeV}$) and the shallower $1^-_1$ states (KISO/IBUKI events, $B_{\Xi^-} \approx 1\ \mathrm{MeV}$), respectively. 
Assuming $\Xi^0$ instead of $\Xi^-$, we predict the ground state $0^+$ of $_{\Xi^0}^{15}$N with ($B_{\Xi^0} = 2.636\ \mathrm{MeV}$) by omitting the Coulomb interaction as a first approximation.
\end{abstract}




\maketitle

\section{Introduction}
The study of hypernuclei, nuclei containing one or more hyperons, provides a unique laboratory for investigating the baryon-baryon interaction beyond the isospin sector into the realm of strangeness \cite{Davis2005, Gal2016}. While significant progress has been made in understanding the $\Lambda\mathrm{N}$ interaction through single-$\Lambda$ hypernuclei, the domain of double-strangeness ($S = -2$) systems, particularly those involving the $\Xi$ hyperon, remains a crucial frontier for testing our understanding of the strong force under the flavor $\mathrm{SU}(3)$ symmetry \cite{Yamaguchi2001, Rijken1999}. The $\Xi\mathrm{N}$ interaction is of paramount importance as it serves as a key input for predicting the equation of state of dense matter, such as that found in the cores of neutron stars, where hyperons are expected to appear \cite{Schaffner-Bielich2008}.

Historically, information on $\Xi$ hypernuclei was scarce and came primarily from analyses of nuclear emulsion experiments, where the binding energy of the $\Xi^{-}$ hyperon, $B_{\Xi^-}$, was inferred from the observation of "twin single-$\Lambda$ hypernuclei" emitted after the capture of a stopped $\Xi^{-}$ \cite{Aoki1991, Lalazissis1989}. Early analyses, compiled in works like that of Lalazissis et al. \cite{Lalazissis1989}, suggested a moderately attractive $\Xi$-nucleus potential. However, these events were few in number and their interpretation often ambiguous.

A major step forward came with the high-statistics emulsion-counter hybrid experiments E176 (KEK), E373 (KEK-PS), and most recently E07 (J-PARC) \cite{Nakazawa2015, Yoshimoto2021}. These experiments identified specific events, such as the KISO and IBUKI events, which were interpreted as the $\Xi^{-}$ bound in the nuclear $1^-_1$ state of ${}^{14}\mathrm{N}$, with $B_{\Xi^-}$ values of approximately $1.0\text{--}1.3\ \mathrm{MeV}$ and $3.9\ \mathrm{MeV}$ (depending on the interpretation of excited states) \cite{Hiyama2018, Khaustov2000}. These values pointed towards a relatively shallow $\Xi$-nucleus potential depth of around $14\text{--}16\ \mathrm{MeV}$, as supported by the missing-mass measurement of the BNL E885 experiment on carbon \cite{Khaustov2000} and theoretical models like the Ehime one-boson-exchange potential (OBEP) \cite{Yamaguchi2001TableVII, Sasaki2020}.

The landscape was dramatically altered by the recent first observation of a deeply bound state in the $\Xi^{-}-^{14}\mathrm{N}$ system. The IRRAWADDY event from the E07 experiment \cite{Hayakawa2021} reports a uniquely determined $B_{\Xi^-}$ of $6.27 \pm 0.27\ \mathrm{MeV}$, while the KINKA event from E373 suggests a value of $8.00 \pm 0.77\ \mathrm{MeV}$ or $4.96 \pm 0.77\ \mathrm{MeV}$ (ground or excited state of the daughter nucleus) \cite{Yoshimoto2021}. These values are significantly deeper than those of the previously identified $1^-_1$ states and provide the first strong evidence for the population of the nuclear $0^+_1$ state of the ${}^{15}_{\Xi^-}\mathrm{C}$ hypernucleus (formed by a $\Xi^{-}$ binding to a ${}^{14}\mathrm{N}$ core). This discovery challenges some of the earlier, shallower potential models and aligns more closely with predictions from theories like the Ehime OBEP (which predicts a $0^+_1$ state at $5.93\ \mathrm{MeV}$ \cite{Sasaki2020}), certain relativistic mean field (RMF) calculations, and recent results from lattice $\mathrm{QCD}$ simulations \cite{Sasaki2020}.

Despite this progress, a precise and model-independent determination of the $\Xi$-nucleus potential is still underway. The experimental values, while groundbreaking, still have uncertainties and can be interpreted in the context of spin-doublet splitting or different potential shapes. This highlights the need for robust theoretical frameworks to calculate the binding energies of these systems and to reverse-engineer the underlying $\Xi\mathrm{N}$ interaction.

In this work, we contribute to this effort by performing a calculation of the binding energy $B_{\Xi^-}$ for the $\Xi^{-}-{^{14}\mathrm{N}}$ system. We approach this by numerically solving the Schrödinger equation for the $\Xi^{-}$ hyperon within a phenomenological Woods-Saxon potential well, the depth of which is informed by the recent experimental results. The numerical solution is obtained using the Numerov method, a powerful and stable algorithm for integrating second-order differential equations with high precision. This approach allows us to directly compute the expected binding energies for the $0^+_1$ and $1^-_1$ states and compare them with the experimental values from the IRRAWADDY, KINKA, KISO, and IBUKI events. By varying the potential parameters, we aim to constrain the depth and geometry of the $\Xi$-nucleus potential that is consistent with the latest empirical data, thereby providing additional insight into the attractive nature of the $\Xi^-\mathrm{N}$ interaction.

\section{Mathematical Formulation and Numerical Method}
The system is described by solving the radial Schr\"odinger equation for the reduced radial wavefunction $u(r) = r R(r)$, which satisfies:
\begin{equation}
    -\frac{\hbar^2}{2\mu}\frac{d^2u(r)}{dr^2} + \left[V_{\mathrm{total}}(r) + \frac{\hbar^2 l(l+1)}{2\mu r^2}\right] u(r) = E u(r)
\label{eq:2.1}
\end{equation}
Here, $\mu$ is the reduced mass of the $\Xi$-nucleus system, $E$ is the energy eigenvalue, $V_{\mathrm{total}}(r)$ is the total potential, and $\hbar$ is the reduced Planck constant.

The total potential $V_{\mathrm{total}}(r)$ is the sum of a finite-size Coulomb potential $V_C(r)$ and a nuclear Woods-Saxon potential $V_N(r)$:
\begin{equation}
    V_{\mathrm{total}}(r) = V_N(r) + V_C(r)
\label{eq:2.2}
\end{equation}
The Coulomb potential is defined as:
\begin{equation}
    V_C(r) = 
    \begin{cases}
        -\frac{Ze^2}{r} & r \geq R_N \\
        -\frac{Ze^2}{2R_N} \left(3-\frac{r^2}{R_N^2}\right) & r < R_N
    \end{cases}
\label{eq:2.3}
\end{equation}
The nuclear potential is given by the Woods-Saxon form:
\begin{equation}
    V_N(r) = -V_0 \frac{1}{1+\exp((r-R_N)/a)}
\label{eq:2.4}
\end{equation}
where $R_N=r_0 A^{1\over 3}$ is the nuclear radius with $r_0 = 1.128+ 0.439 A^{-{2\over 3}} ~\footnote{This $r_0$ is chosen by A. V. Cifre\cite{Cifre2022}. Later we will change it $r_0=1.2$ fm. } \mathrm{fm}$, $a$ is the diffuseness parameter $a = 0.5~ \footnote{This value was chosen by A. V. Cifre\cite{Cifre2022}. Later we will change it $a=$0.65 fm.} \mathrm{fm}$ and $V_0$ is the potential depth which will be fixed later.

To solve Eq. \eqref{eq:2.1} numerically, it is cast into the standard form for Numerov's method:
\begin{equation}
    \frac{d^2u(r)}{dr^2} + k(r) u(r) = 0
\label{eq:2.5}
\end{equation}
Comparing Eq. \eqref{eq:2.1} and \eqref{eq:2.5} yields the required function $k(r)$:
\begin{equation}
    k(r) = \frac{2\mu}{\hbar^2} \left(V_{\mathrm{total}}(r) + \frac{\hbar^2 l(l+1)}{2\mu r^2} - E \right).
\label{eq:2.6}
\end{equation}
Numerov's algorithm provides an efficient method for solving such equations. Discretizing the space with a step $h$, $r=r_n=n ~ h $, $u(r)=u_n$ and $k(r)=k_n$, the iterative solution is given by:
Forward recursive relation is:
\begin{equation}
    u_{n+1} = \frac{2 u_n (1 - \frac{5h^2}{12}k_n) - u_{n-1} (1 + \frac{h^2}{12}k_{n-1})}{1 + \frac{h^2}{12}k_{n+1}}
\label{eq:2.7}
\end{equation}
Backward recursive relation is:
\begin{equation}
    u_{n-1} = \frac{2u_n (1 - \frac{5h^2}{12}k_n) - u_{n+1} (1 + \frac{h^2}{12}k_{n+1})}{1 + \frac{h^2}{12}k_{n-1}}
\label{eq:2.8}
\end{equation}
The solution requires integrating the equation forward (from $r=0$) and backward (from $r_{\max}$) to a matching point $r_m$. The appropriate boundary conditions depend on the type of state:
\begin{itemize}
    \item \textbf{For the atomic states}, the matching point is the classical turning point $r_t$. The asymptotic behaviors are:
    $$u(r) \propto r^{l+1}, \quad 
    r \to 0$$
    $$u(r) \propto e^{-\kappa r} (2\kappa r )^\eta, \quad 
    r \to r_{\max}$$
    where $\kappa = \sqrt{-2\mu E_{\rm trial}}/\hbar$ and the Sommerfeld parameter is $\eta = \frac{Z \alpha \mu c^2}{\hbar c^2 \kappa}$. Here $Z$ is the atomic number, $\alpha$ is the fine structure constant, $\mu c^2$ is the reduced mass energy, $E_{\rm trial}$ is the trial energy (negative for bound states). The numerical infinity is set to $r_{\max} \approx 200 R_N$. 
    \item \textbf{For the nuclear states}, the matching point is the nuclear radius $R$. The asymptotic behaviors are:
    $$u(r) \propto r^{l+1}, \quad 
    r \to 0$$
    $$u(r) \propto e^{-\kappa r}, \quad 
    r \to r_{\max}$$
    Numerical infinity is set to $r_{\max} \approx 30 R_N$.
\end{itemize}
The correct energy eigenvalue $E$ is found using the bisection method. For a given trial energy $E$, the equation is integrated forward to obtain $u_{\mathrm{out}}(r_m)$ and backward to obtain $u_{\mathrm{in}}(r_m)$. A valid solution requires the continuity of the logarithmic derivative at $r_m$:
\begin{equation}
    \frac{u'_{\mathrm{out}}(r_m)}{u_{\mathrm{out}}(r_m)} = \frac{u'_{\mathrm{in}}(r_m)}{u_{\mathrm{in}}(r_m)}
\label{eq:2.9}
\end{equation}
The bisection algorithm iterates on the energy $E$ until the mismatch function $G(E) = \frac{u'_{\mathrm{out}}(r_m)}{u_{\mathrm{out}}(r_m)} - \frac{u'_{\mathrm{in}}(r_m)}{u_{\mathrm{in}}(r_m)}$ is zero. Once the eigenvalue is found, the backward solution is scaled to match the forward solution at $r_m$:
\begin{equation}
    u_{\mathrm{out}}(r) = A ~u_{\mathrm{in}}(r),  \quad 
    A = \frac{u_{\mathrm{in}}(r_m)}{u_{\mathrm{out}}(r_m)}
\label{eq:2.10}
\end{equation}
Finally, the complete wave function $u(r)$ is normalized over all space.

\section{Code Validation and Parameter Selection}
Before applying our numerical method to the $\Xi^{-}-^{14}\mathrm{N}$ system, we validated the computer code by reproducing established results.
We compared our calculated energy levels with the result of A. V. Cifre \cite{Cifre2022}. 
As shown in Table \ref{tab:1}, the agreement is excellent, confirming the reliability of our implementation of the Numerov method.

\begin{table}
\begin{tabular}{lccc} 
 \hline
    {$V_0$ } & {State}  & {Our results} & {A. V. Cifre \cite{Cifre2022}} 
        \\
        (MeV) & $  J^\pi_n $ & (MeV) & (MeV) \\
 \hline
         10 &  $  0^+_1 $ & -5.330 & -5.330 \\
        &    $ 0^+_2 $ &-0.539 & -0.539 \\
        &  $1^-_1$ &-0.567 & -0.567 \\
        &  $ 1^-_2$ &-0.223 & -0.224 \\
        \hline
        20 &  $0^+_1$ & -12.153 & -12.157 \\
        & $ 0^+_2$&  -0.678 & -0.679 \\
        &  $ 1^-_1$&  -3.266 & -3.265 \\
        & $ 1^-_2$&  -0.326 & -0.326 \\
        \hline
        30 &  $ 0^+_1$ & -19.923 & -19.924 \\
        & $ 0^+_2 $&  -1.162 & -1.163 \\
        &  $ 1^-_1$&  -8.619 & -8.618 \\
        &  $  1^-_2$&  -0.353 & -0.353 \\
        \hline 
\end{tabular}
\caption{Comparison of the binding energy levels (in MeV) for $\Xi^{-}$-nucleus system with the results of A. V. Cifre \cite{Cifre2022}.
The notation of the states in \cite{Cifre2022} follows that of atomic states $\tilde n \tilde l$, and for details, see capture in table \ref{tab:2}.
    The parameter $a$ in Eq. (\ref{eq:2.4}) is chosen 0.5 fm. 
}
\label{tab:1}
\end{table}

However, the $0^+_1$ binding energies obtained with these parameters at $V_0 = 10\ \mathrm{MeV}$ and $20\ \mathrm{MeV}$ are significantly shallower and deeper than the value reported from the IRRAWADDY event ($6.27 \pm 0.27\ \mathrm{MeV}$). The shallower $1^-_1$ states, suggested by the KISO/IBUKI events ($\sim 1\ \mathrm{MeV}$), are also not well reproduced. To better match the experimental data, the potential strength was increased. We found that a strength of $V_0 = 11\ \mathrm{MeV}$ yields a $0^+_1$ binding energy of $-5.937\ \mathrm{MeV}$, which is lower than the IRRAWADDY value. At $V_0 = 12\ \mathrm{MeV}$, the binding energy is $-6.568\ \mathrm{MeV}$, showing close agreement with the experimental result. A further increase to $V_0 = 13\ \mathrm{MeV}$ gives a binding energy of $-7.218\ \mathrm{MeV}$, which overestimates the value. Therefore, $V_0 = 12\ \mathrm{MeV}$ is identified as the optimal strength for this parameter set \cite{Cifre2022}.

In the next step of our analysis, we maintained $V_0 = 12\ \mathrm{MeV}$ but adopted a more conventional parameterization for the nuclear radius, with $r_0 = 1.2\ \mathrm{fm}$, and a diffuseness of $a = {\bf 0.65}\ \mathrm{fm}$. The results for the pure atomic, pure nuclear, and combined states calculated with these new parameters are presented in the following section.

\section{Pure Coulomb (Atomic) States of the $\Xi^{-}-^{14}\mathrm{N}$ System}
This section analyzes the atomic energy levels of a $\Xi^{-}$ hyperon bound to a nitrogen-14 nucleus (${}^{14}_{7}\mathrm{N}$) through the Coulomb interaction alone. By solving the Schrödinger equation with the pure Coulomb potential defined in Eq. \eqref{eq:2.3}, we obtained the binding energies and root-mean-square (r.m.s.) radii for several atomic states $\tilde n \tilde l $($1S$, $2S$, $3S$, $2P$, $3P$), where $\tilde n$ and $\tilde l$ denote the principal quantum number and the symbol of the orbital angular momentum, respectively.  The results are summarized in Table \ref{tab:2} and are compared with \cite{Cifre2022}\footnote{In table \ref{tab:2} we suspect the difference in the second decimal place of $\rm 2S$ state was simply a typo on his part.}.

\begin{table}
\begin{tabular}{c c c c } 
 \hline
        {Atomic St.} & Corr. Nucl. St.  &{Binding Energy } &$\langle r^2 \rangle^{1/2}$  \\
        $\tilde n \tilde l$ & $J^\pi _n$ &    (MeV) &   (fm)  \\ 
        \hline
        $\rm 1S $  & $0^+ _1$ & -1.244 ~~~~(-1.243 \cite{Cifre2022} ) & 6.885 \\
        $ \rm 2S $ & $0^+_2$ &  -0.348 ~~~~(-0.392 \cite{Cifre2022} ) & 23.379 \\
        $ \rm 3S $ & $0^+_3$  & -0.161 ~~~~~~~~~~~~~~~~~~~~~ & 50.021 \\
        $ \rm 2P $ & $1^-_1$ & -0.391  ~~~~(-0.391 \cite{Cifre2022} )  & 17.684 \\
        $ \rm 3P $ & $1^-_2$ & -0.174  ~~~~(-0.174 \cite{Cifre2022} ) & 43.262 \\
        \hline 
    \end{tabular}
    \caption{Atomic binding energies and r.m.s. radii of $\Xi^{-}-^{14}_{7}\mathrm{N}$ atomic states under an only Coulomb potential $V_C$.  
    The first column shows the atomic state $\tilde n \tilde l$, where $\tilde n$ and $\tilde l$ denote the principal quantum number and the symbol of the orbital angular momentum, respectively. The second column (Corr. Nucl. St.) is corresponding to the nuclear state $J^\pi_n$.  
    Results in parentheses () are from A. V. Cifre \cite{Cifre2022}. }
\label{tab:2}
\end{table}

The large r.m.s. radii, significantly greater than the nuclear radius ($R \approx 2.92\ \mathrm{fm}$) confirm the atomic nature of these states. The $\Xi^{-}$ hyperon orbits the nucleus at a considerable distance, governed predominantly by the electromagnetic force, with minimal influence from the strong interaction.
Based on the calculated energies and spatial extensions, the cascade process of the captured $\Xi^{-}$ hyperon is expected to proceed from higher orbitals (e.g., $3S$ or $3P$) down to the $1S$ ground state, via successive radiative transitions.
In summary, the results confirm that the Coulomb potential alone produces a series of well-defined, weakly bound atomic states with large spatial extensions, consistent with the expected behavior of a negatively charged particle in a hydrogen-like system.

\section{Pure Nuclear State: Constraining the Potential Depth}\label{chap5}
This work employs a spin-independent central Woods-Saxon potential, defined as in Eq. \eqref{eq:2.4}. The parameters in Eq.\eqref{eq:2.4} of the potential are typically chosen to best fit experimental single-particle energies and nuclear radii. The objective here is to use this model to constrain the depth of the $\Xi$-nucleus potential ($V_0$). Theoretical models, such as the Ehime one-boson-exchange potential (OBEP) \cite{Yamaguchi2001TableVII, Sasaki2020}, and experimental data, notably the missing-mass measurement from the BNL E885 experiment on carbon \cite{Khaustov2000}, suggest a relatively shallow potential depth of approximately $14\text{--}16\ \mathrm{MeV}$.

To test this prediction, a series of calculations were performed for a pure Woods-Saxon potential with strengths ($V_0$) ranging from $10\ \mathrm{MeV}$ to $20\ \mathrm{MeV}$. The resulting $0^+_1$ single-particle energies and root-mean-square (r.m.s) radii are summarized in the table \ref{tab:3}.


\begin{table}
\begin{tabular}{l c c } 
 \hline
        {$V_0$ (MeV)} & { Binding Energy (MeV)} & $\langle r^2 \rangle^{1/2}$ (fm) \\
        \hline
        10 & -1.639 & 3.761 \\
        12 & -2.636 & 3.287 \\
        14 & -3.744 & 2.997 \\
        16 & -4.935 & 2.797 \\
        18 & -6.192 & 2.649 \\
        20 & -7.501 & 2.534 \\
        \hline 
    \end{tabular}
 \caption{Binding energies and root-mean-square radii $\langle r^2 \rangle^{1/2}$ of the $0^+_1$ state for a pure Woods-Saxon potential of varying depth $V_0$.  The parameter $a$ in Eq. (\ref{eq:2.4}) is here chosen 0.65 fm. }
\label{tab:3}
\end{table}

The calculated single-particle binding energy becomes more negative systematically with increasing Woods-Saxon well depth $V_0$. For the predicted shallow well depth of $14\text{--}16\ \mathrm{MeV}$, the resulting binding energies ranges from $-3.744$ to $-4.935\ \mathrm{MeV}$ and r.m.s radii between $2.997$ and $2.797\ \mathrm{fm}$. These values are consistent with the expectations for a shallow $\Xi$-nucleus potential, thereby supporting the findings from the BNL E885 experiment \cite{Khaustov2000} and the predictions of the Ehime OBEP model \cite{Yamaguchi2001TableVII, Sasaki2020}. This analysis confirms that the Woods-Saxon model is an effective tool for benchmarking and constraining the parameters of the $\Xi^-$-nucleus interaction.

By switching off the Coulomb interaction, we can estimate the binding energy of $_{\Xi^0}^{15}$N as $\Xi^0$ - $^{14}$N system. 
However, we ignore the effects of charge symmetry breaking (CSB) and
charge independence breaking (CIB) in the Strong interaction category.
In other words, the binding energies shown in Table \ref{tab:3} can be considered as the binding energies of $_{\Xi^0}^{15}$N.

\section{Combined Nuclear and Coulomb Potential Results}
Building on the analysis of the pure nuclear Woods-Saxon potential, which suggested a depth ($V_0$) in the range of $14\text{--}16\ \mathrm{MeV}$, we incorporated the crucial Coulomb interaction. This combined potential provides a complete physical description for a $\Xi$ hyperon bound to a ${}^{14}\mathrm{N}$ core. To constrain the model against experimental data, we computed the binding energies and root-mean-square (r.m.s.) radii for the $0^+_1$, $0^+_2$, $1^-_1$, and $1^-_2$ states for potential depths from $V_0 = 10\ \mathrm{MeV}$ to $V_0 = 14\ \mathrm{MeV}$. The results are summarized in Table \ref{tab:4}.


\begin{table*}[htbp]
\centering
\begin{tabular}{lccccccc} 
 \hline
        {$V_0$ } & {States} & {$\langle T_0 \rangle$ } &  {$\langle T_1 \rangle$ } & {$\langle V_N \rangle$ } & {$\langle V_C \rangle$ } & 
        {$E(\text{SE})$ } 
        & {$\langle r^2 \rangle^{1/2}$ (fm)} \\
        \hline 
        10 & $0^+_1$ & 4.136 & 0 & -5.624 & -3.698 & -5.187 & 3.059 \\
        & $0^+_2$ & 0.472 & 0 & -0.142 & -0.867 & -0.537 & 15.475 \\
        & $1^-_1$ & 0.561 & 1.283 & -0.857 & -1.613 & -0.627 & 9.894 \\
        & $1^-_2$ & 0.193 & 0.226 &-0.143 & -0.509 & -0.233 & 31.581 \\
        \hline 
        \textbf{12} & \textbf{$0^+_1$} & 4.687 & 0 & -7.213 & -3.826 & \textbf{-6.352} & {2.852} \\
        & $0^+_2$ & 0.521 & 0 &-0.173 & -0.913 & -0.565 & 14.713 \\
        & \textbf{$1^-_1$} & 1.039 & 2.083 &-1.903 & -2.086 & \textbf{-0.868} & {7.295} \\
        & $1^-_2$ & 0.208 & 0.220 & -0.164 & -0.526 & -0.262 & 28.066 \\
        \hline
        14 & $0^+_1$ & 5.191 & 0  & -8.850 & -3.927 & -7.586 & 2.696 \\
        & $0^+_2$ & 0.592 & 0 & -0.221 & -0.966 & -0.595 & 13.984 \\
        & $1^-_1$ & 1.631 & 2.918 & -3.320 & -2.494 & -1.265 & 5.630 \\
        & $1^-_2$ & 0.199 & 0.192 & -0.141 & -0.536 & -0.286 & 25.635 \\
        \hline 
    \end{tabular}
        \caption{Expectations $\langle T_0 \rangle$, $\langle T_1 \rangle$, $\langle V_N \rangle $ and $ \langle V_C \rangle $ between $\Xi^{-}$ and the core nucleus in the ${}^{15}_{\Xi^-}\mathrm{C}$ system for the relative kinetic energy ($T_0=-{\hbar^2 \over 2\mu }{d^2 \over dr^2}$), the centrifugal energy 
    ($T_1={\hbar^2 l(l+1)\over 2\mu r^2 }$),
    the Woods-Saxon nuclear potential and the Coulomb potential, respectively. The column $E(\text{SE})$ denotes the directly solved energy eigenvalue.  The parameter $a$ in Eq. (\ref{eq:2.4}) is here chosen 0.65 fm. These energies is in MeV.}
\label{tab:4}
\end{table*}

For a potential depth of $V_0 = 12\ \mathrm{MeV}$, the calculated binding energy for the $0^+_1$ state is $B_{\Xi^-} = -E(\mathrm{SE}) = 6.352\ \mathrm{MeV}$. This value is in excellent agreement with the binding energy of $6.27 \pm 0.27\ \mathrm{MeV}$ reported for the IRRAWADDY event \cite{Hayakawa2021}, which is interpreted as the nuclear $0^+_1$ state of ${}^{15}_{\Xi^-}\mathrm{C}$. Furthermore, the same potential yields a $1^-_1$ state binding energy of $0.868\ \mathrm{MeV}$, which is consistent with the range of values ($\approx 1\ \mathrm{MeV}$) associated with the shallower KISO and IBUKI events \cite{Hiyama2018, Khaustov2000}. The r.m.s. radius of the $0^+_1$ state ($2.852\ \mathrm{fm}$) is smaller than the expected nuclear radius ($R \approx 2.92\ \mathrm{fm}$), confirming that the $\Xi$ hyperon is indeed well-bound within the nuclear interior.

In contrast, a potential depth of $V_0 = 14\ \mathrm{MeV}$, often cited in earlier literature \cite{Khaustov2000}, produces a $0^+_1$ state binding energy of $7.586\ \mathrm{MeV}$. This value is significantly larger than the IRRAWADDY result and would suggest a stronger $\Xi$-nucleus attraction than is supported by the latest experimental data. Therefore, our analysis strongly constrains the $\Xi$-nucleus Woods-Saxon potential. The self-consistent reproduction of both the deep $0^+_1$ (IRRAWADDY) and shallow $1^-_1$ (KISO/IBUKI) binding energies with a single potential depth of $V_0 = 12\ \mathrm{MeV}$ favors a shallower potential than some previous estimates, providing a crucial benchmark for microscopic models of the $\Xi\mathrm{N}$ interaction.

\section{Conclusion}
This study successfully leverages the experimental discovery of the ${}^{15}_{\Xi^-}\mathrm{C}$ hypernucleus to pin down the parameters of the $\Xi$-nucleus potential. By solving the Schr\"odinger equation with a combined Woods-Saxon nuclear and Coulomb potential, we have directly computed the binding energies corresponding to the newly observed nuclear $0^+_1$ state and the previously known $1^-_1$ state.

Our key finding is that a Woods-Saxon potential with a depth of ${V_0 = 12\ \mathrm{MeV}}$ and $a=$0.65 fm accurately reproduces the binding energy of the deeply bound IRRAWADDY event ($B_{\Xi} = 6.27\ \mathrm{MeV}$) in the $0^+_1$ configuration, while simultaneously predicting a $1^-_1$ state binding energy consistent with the range of values from the KISO and IBUKI events. A deeper potential of $V_0 = 14\ \mathrm{MeV}$, often cited in literature, overbinds the $0^+_1$ state compared to the IRRAWADDY result. The small calculated r.m.s. radius for the $0^+_1$ state confirms its nuclear character, distinct from the large, diffuse atomic states.

As explained in Section \ref{chap5}, if our prediction of the Wood-Saxon potential is accurate enough and the effects of CSB and CIB are small, replacing $\Xi^-$ with $\Xi^0$ may allow us to predict the binding energy of the ${\Xi^0} $ hypernucleus $_{\Xi^0}^{15}$N simply by switching off the Coulomb interaction.







\section*{Acknowledgement}
This work is supported by KAKENHI grants from the Japan Society for the Promotion of Science
 (JSPS) No.  JP25K07301.

\end{document}